# Designing biodegradable surfactants and effective biomolecules with dissipative particle dynamics

Armando Gama Goicochea

Departamento de Ciencias Naturales, DCNI, Universidad Autónoma Metropolitana Unidad Cuajimalpa, Av. Pedro Antonio de los Santos 84, México, D. F. 11850, Mexico.

E-mail: agama@alumni.stanford.edu

## Abstract

The design of a biodegradable, environmentally friendly surfactant is carried out, taking the structure of a known surfactant that lacks these qualities as the starting point, using mesoscopic computer simulations. The newly designed surfactant is found to perform at least as well as its predecessor, without the latter´s inimical characteristics. In the second part of this work, a comparative study of model proteins with different amino acid sequence interacting with surfaces is undertaken. The results show that, all other aspects being equal, this sequence is the key factor determining the optimal activity of the proteins near surfaces. These conclusions are found to be in agreement with recent experiments from the literature.

## 1 Introduction

Amphiphilic molecules such as surfactants, polymers and proteins are very important in biological processes such as drug delivery, adsorption on living tissue, and molecular association (Jönsson et al., 1998). Although the basic interactions between the various molecules are often well known (van der Waals, electrostatic), the complex interplay that emerges from the many – body manifestations of these forces with the specific structures of the amphiphilic molecules is not. Detailed knowledge of how these interactions play in complex fluids composed of those molecules, biological membranes and water is not easily accessible from a theoretical point of view because it represents a scenario with vastly different length and time scales (Israelachvili, 2011). Fortunately, the recent advances in the speed of microprocessors have made it possible to solve computer models of biologically relevant systems in reasonable times, thus providing important information to understand, improve and design new biomolecules tailored to solve specific needs.

The work presented here reports the importance of the structure of biomolecules in their performance in environments of current interest, by means of mesoscopic, dissipative particle dynamics (DPD) computer simulations. The first part is devoted to showing how an environmentally friendly non ionic surfactant was designed starting from one that was not, without detriment to its performance. The prediction of the thermodynamic properties of the new surfactant led to the conclusion that they were at least of equal quality as those of its predecessor, with the structure of the surfactants playing the major role. In the second part, I show how a model protein with different amino acid sequence can lead to entirely different thermodynamic conditions when placed near a lipid membrane, which defines a criterion for choosing the best one before synthesizing one in the lab. The underlying thesis of this contribution is that, all things being equal, the structure of the molecules defines their function in a complex biological environment.

The rest of this chapter is organized as follows. In Section 2 I introduce the basics of the DPD model and simulation details. The next section is devoted to the presentation of a newly designed, environmentally friendly and biodegradable surfactant and the comparison of its performance with a commercially available (not environmentally friendly), similar surfactant. Section 4 is dedicated to the study of the influence that the amino acid sequence has on the thermodynamic behaviour of model proteins interacting with biologically relevant surfaces. The conclusions are drawn in Section 5.

## 2 The DPD model

Atomistically detailed computer simulations (Allen and Tildesley, 1987) have proved to be very successful, but in order to achieve scales comparable to those accessible to experiments they require considerable computational resources. When modelling large molecules and long time scales as it is befitting in biological systems, one needs tools that go beyond the atomistic regime, and DPD (Hoogerbrugge and Koelman, 1992) is one of them. The reason relies on the fact that DPD involves short – range, linearly decaying forces which can be integrated using a time step that is at least three orders of magnitude larger than that used in microscopic simulations, allowing the study of phenomena at the mesoscopic scale.

The basic structure of DPD consists of three forces, conservative $F_{ij}^C$, dissipative $F_{ij}^D$ and random $F_{ij}^R$. All forces between particles $i$ and $j$ have simple spatial dependences and vanish beyond a finite cutoff radius $R_c$, which represents the intrinsic length scale of the DPD model and it is usually chosen as the reduced unit of length, $R_c = 1$. Their functional expressions are:

$$F_{ij}^C = a_{ij}\left(1 - {r_{ij}}/{R_c}\right)\hat{e}_{ij} \tag{1}$$

$$F_{ij}^D = -\gamma\left(1 - {r_{ij}}/{R_c}\right)^2 [\hat{e}_{ij}\cdot v_{ij}]\hat{e}_{ij} \tag{2}$$

$$F_{ij}^R = \sigma\left(1 - {r_{ij}}/{R_c}\right)\hat{e}_{ij}\xi_{ij}. \tag{3}$$

In the expressions above, $r_{ij} = r_i - r_j$ is the relative position vector, $\hat{e}_{ij}$ is the unit vector in the direction of $r_{ij}$, and $v_{ij} = v_i - v_j$ is the relative velocity, with $r_i, v_i$ the position and velocity of particle $i$, respectively. The variable $\xi_{ij}$ is generated randomly between 0 and 1 with a Gaussian distribution of unit variance; $a_{ij}$, $\gamma$ and $\sigma$ are the strength of the conservative, dissipative and random forces, respectively; all forces are zero for $r_{ij} > R_c$. All beads are the same size, but the difference between beads of different chemical nature is defined by the constant $a_{ij}$, and all thermodynamic properties are dependent on it. The strengths of the random and dissipative forces are related as follows:

$$\frac{\sigma^2}{2\gamma} = k_B T \qquad\qquad (4)$$

which represents the fulfilment of the fluctuation – dissipation theorem for DPD (Español and Warren, 1997) and it is one of the defining qualities of the method. Another key feature is that the forces in equations (1) – (3) are pairwise additive, therefore local and global momentum is conserved, which in turn means that all hydrodynamic modes present in the fluid shall be preserved. The conservative force parameter for particles of the same type, $a_{ii}$, is given by $a_{ii} = [(\kappa^{-1}N_m - 1)/2\alpha\rho]k_B T$, where $N_m$ is the coarse-graining degree (number of water molecules grouped in a DPD particle), $\alpha$ is a numerical constant equal to 0.101, $\rho$ is the DPD number density, $\kappa^{-1}$ is the inverse compressibility of the water at room temperature. I choose a coarse-graining degree equal to 3 water molecules in a DPD particle and use the dimensionless water compressibility at standard conditions, $\kappa^{-1} \approx 16$, so that the parameter above is $a_{ii} = 78.3$. The parameter for different types of particles, $a_{ij}$, is calculated from its Flory-Huggins coefficient $\chi_{ij}$ using the relation $a_{ij} = a_{ii} + 3.27\chi_{ij}$. For further details, see (Groot and Warren, 1997). The DPD method has enjoyed considerable success over a wide range of applications, including biological systems. Some of the most recent ones have been reviewed extensively by (Murtola et al., 2009).

## 3 Modelling of a biodegradable surfactant

The ecological impact of surfactants has become increasingly important in most contemporary formulations. The rate at which surfactants will biodegrade at some sewer plant determines to a large extent the preference for one or another surfactant. Some of the aspects that are relevant when monitoring surfactants environmental impact are aquatic toxicity in fish mainly; bioaccumulation resulting from the built up of organic compounds in fish also, and biodegradability. The latter results typically from a series of enzymatic reactions that break down the original composition of the surfactant and turn it into a mix of water, oxides and other by products (Jönsson et al., 1998).

Among the most frequently used surfactants in modern day applications are the non ionic, nonylphenol ethoxylates which are used in emulsion polymerization processes, as detergents and pesticides to name a few. In this section I shall be primarily interested in a particular example of this class of surfactants, whose structure is shown in Figure 1.

$$CH_3(CH_2CH_2)_4\text{-}C_6H_4\text{-}O(CH_2CH_2O)_{10}H$$

**Figure 1** Chemical structure of the surfactant modelled in this work. It is a nonylphenol ethoxylate with 10 moles of ethylene oxide.

Surfactants of the type shown in Figure 1 are useful for many applications but in recent years their use has been considered somewhat deleterious for the environment due to the fact that a linear molecule, such as the one shown in Figure 1, is easily biodegradable. Under typical circumstances this would be a favourable aspect, except for the fact that the surfactant shown in the figure above, which shall be called NP10 (meaning nonylphenol with 10 moles of ethylene oxide) in what follows, contains a benzene ring in its structure. The release of free benzene may lead to the disruption of the balance of hormones in fish and other organisms (Boogaard P. J., van Sittert N. J., (1995)), although it appears to be a not too strong effect. Nevertheless, this is an aspect that deserves attention and as such, in this section I shall study the adsorption properties of NP10 in a model biological environment.

| DPD Particle | Atoms in DPD particle |
|:---:|:---:|
| C | $OCH_2CH_2O$ |
| D | $CH_2CH_2OCH_2$ |
| E | $HOCH_2CH_2$ |
| F | $C_6H_4$ |
| G | $CH_2CH_2CH_2$ |
| H | $CH_3CH_2CH_2$ |

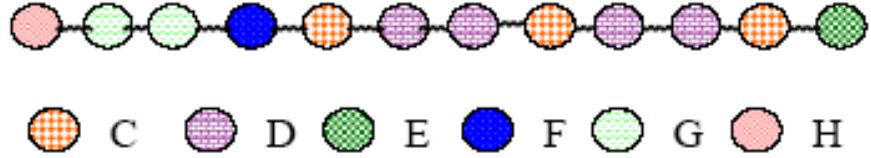

**Figure 2** DPD mapping of the surfactant. Each bead represents a group of atoms. The top part of the figure shows the specific atomic grouping associated with each DPD bead shown in the bottom part of this figure.

To proceed, the first step is to map the structure of NP10 shown in Figure 1 to DPD beads. Then, a model for the adsorbing surfaces must be introduced. Following the standard procedure (Groot and Warren, 1997) for a coarse graining degree equal to 3 water molecules per DPD bead, one obtains the coarse – graining mapping of surfactant NP10 as shown in Figure 2. The matrix of conservative interaction parameters $a_{ij}$ (see equation 1) obtained after applying such procedure is presented in Table 1, where the interaction parameters of each DPD particle of the NP10 surfactant (shown in Figure 2) with a model biological membrane, $a_{iw}$, are listed as well.

| $a_{ij}$ | 1 | 2 | 3 | 4 | 5 | 6 | 7 | $a_{iw}$ |
|---|---|---|---|---|---|---|---|---|
| 1 | 78.3 | 39.15 | 39.15 | 39.15 | 155 | 171.8 | 167.5 | 31.2 |
| 2 | | 78.3 | 79.3 | 86.3 | 79 | 81.3 | 80.5 | 31.5 |
| 3 | | | 78.3 | 92.9 | 78.3 | 78.8 | 78.6 | 31.5 |
| 4 | | | | 78.3 | 91.5 | 99 | 97 | 31.5 |
| 5 | | | | | 78.3 | 79.1 | 78.8 | 17.5 |
| 6 | | | | | | 78.3 | 78.4 | 21 |
| 7 | | | | | | | 78.3 | 21 |

**Table 1** Conservative DPD interaction parameters ($a_{ij}$) used in the simulations of adsorption of NP10. The numbers identify each type of particle in the simulations, starting with water (1), and the surfactant DPD particles defined in Figure 2: C (2), D (3), E (4), F (5), G (6) and H (7). The last column lists the interaction ($a_{iw}$) of each type of DPD particle with the surfaces, see equation 5.

To model a soft biological membrane on which NP10 will adsorb, the following short range, effective force is added:

$$\boldsymbol{F_{iw}}(z_i) = a_{iw}\left(1 - {z_i}/{R_c}\right)\boldsymbol{\hat{e}_k}.\qquad(5)$$

In equation 5, the force between a particle $i$ (either water or the NP10 surfactant) and the wall representing a smooth biomembrane placed parallel to the $XY$ – plane, at the ends of the simulation box in the $z$ – direction, is shown to depend only on the component of such particle's position along the $Z$ – axis. The wall interaction constant $a_{iw}$, whose values for the various DPD particles in the simulation box have been listed in Table 1, and they were obtained following the same procedure as that used for the fluid's particle – particle interactions. The symbol $\boldsymbol{\hat{e}_k}$ represents the unit vector in the direction perpendicular to the $XY$ – plane. Equation 5 represent a soft surface, in much the same spirit as the soft interactions given by equation 1, as it is appropriate for example, for surfaces formed of lipid bilayers (Israelachvili, 2011). More sophisticated, self – consistent DPD surface models are available (Gama Goicochea and Alarcón, 2011), but they are better suited to study harder solid walls, which are not of biological relevance, within the context of the present study.

For the modeling of the adsorption of NP10 on the membranes the following procedure is followed. A given concentration of NP10 is chosen and allowed to reach equilibrium while keeping the chemical potential of the solvent constant. Once equilibrium has been reached, one determines the amount of NP10 that was adsorbed and what was left in the supernatant, i.e. not adsorbed, if any. This procedure is followed for as many surfactant concentrations as one wishes, constructing the adsorption isotherm from the data collected. Experimental adsorption isotherms are obtained following precisely the same method (Kronberg, 2001). Monte Carlo (MC) simulations in the Grand Canonical (GC) statistical mechanical ensemble (which fixes the chemical potential, $\mu$, as well as the volume and temperature) were carried out to obtain the equilibrium state for each NP10 concentration, using a code hybridized with DPD, see (Gama Goicochea, 2007) for full details, including the integration algorithm for the equation of motion, with a time step $\delta t$=0.03. Only the number of solvent particles was allowed to fluctuate, to keep the chemical potential fixed. The dimensions of the simulation box were fixed at $L_x$=7, $L_y$=7 and $L_z$=14 adimensional DPD units. The temperature was fixed at $T$=1 by choosing $\sigma$=3 and $\gamma$=4.5 (see equations 2 and 3), and the solvent's chemical was chosen as $\mu$=37.7, which leads to an average total density $\rho$=3; by doing so one assures that the results are invariant under changes of conservative force interaction parameters (Groot and Warren, 1997). For the equilibrium phase, 30 blocks of $10^4$ MC steps were run, followed by an additional $100 \times 10^4$ MC steps for the production phase. The solvent is modeled as monomeric DPD particles, while the surfactant is made up of 12 DPD units, with the sequence as shown in Figure 2, and with the surfactant beads joined by freely rotating harmonic springs with spring constant $k_0=100$ and equilibrium length $r_0=0.7$ (Gama Goicochea, 2007).

The NP10 concentration was varied in the range of $4 \leq [c] \leq 90$ surfactant molecules in the simulation box, with the number of solvent monomers being adjusted by the GCMC ensemble to keep the chemical potential always fixed. At each concentration the density profile of the monomers that make up NP10, $\rho(z)$, was computed, and the amount of adsorbed NP10, $\Gamma$, was obtained from the equation (Gama Goicochea, 2007):

$$\Gamma = \int_0^{L_z} [\rho(z) - \rho_B] \, dz. \qquad (6)$$

In equation 6, $\rho_B$ is the density of NP10 monomers in the bulk, namely those that were not adsorbed on the confining surfaces. If all NP10 molecules were adsorbed, $\rho_B = 0$. Figure 3 shows typical equilibrium configurations for two NP10 concentrations, [c] = 14 molecules/volume (left image), and [c] = 30 molecules/volume.

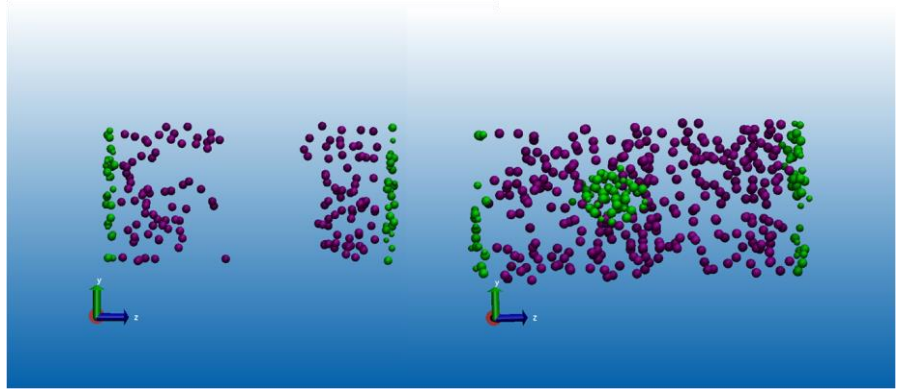

**Figure 3** Adsorption of NP10 on model biomembranes. Green beads represent DPD beads named F, G and H (see mapping in upper panel of Figure 2). Purple beads are hydrophilic surfactant beads, namely C, D, and E. The solvent beads have been removed for clarity. The left image corresponds to a NP10 concentration [c] = 14 molecules/volume, while the one on the right is for [c] = 30. Notice the incipient formation of a micelle in the latter.

As is evident from Figure 3, the surfactant is subjected to two competing factors: on the one hand it has a strong tendency to adsorb at the interfaces (see left image in this figure), especially at relatively low concentrations. But, as the concentration is increased some surfactants find it more favorable to form micelles (right image in Figure 3) while others are adsorbed. In the particular example shown in Figure 3, those NP10 molecules forming the micelle (right image) would be the ones that contribute to the bulk surfactant density, $\rho_B$, in equation 6, while those adsorbed would be accounted for in $\rho(z)$. It is also of notice that the surfactant appears to adsorb in monolayers, in much the same way as assumed by the Langmuir adsorption model (Kronberg, 2001).

From the series of simulations previously described the adsorption isotherm of the NP10 surfactant in water was obtained, and it is presented in Figure 4. It is clearly a Langmuir – type isotherm, from which one can easily obtain the saturation concentration, i. e. the amount of surfactant that has saturated the available adsorption sites on the surfaces, so that whatever additional amount of surfactant added to the system is not likely to be adsorbed. This is an important quantity because it is obtained from experiments. therefore it represents a basis for comparison.

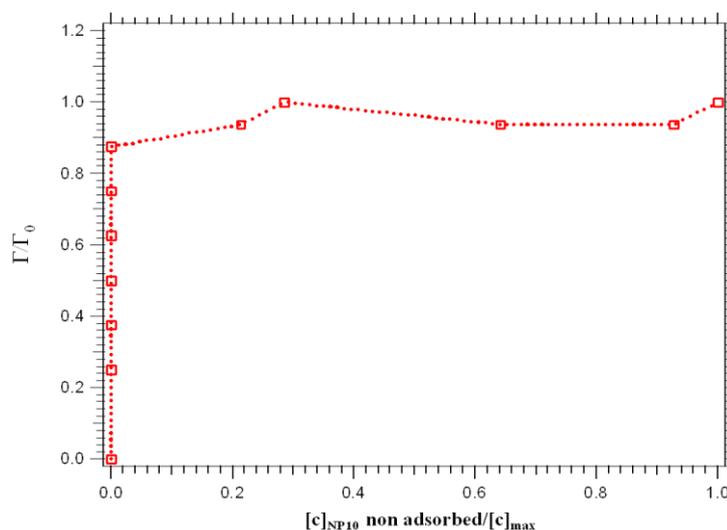

**Figure 4** Adsorption isotherm of NP10 on model biomembranes. The axes have been normalized to allow for comparisons with other surfactants. The $X$ – axis is the number of non adsorbed NP10 molecules, normalized by the maximum

amount of non adsorbed surfactant, $[c]_{max}$. The $Y$ – axis represents the number of adsorbed NP10 molecules for each concentration added to the dispersion, normalized by the maximum adsorbed amount, at saturation, $\Gamma_0$.

The curve in Figure 4 is a typical example of a Langmuir isotherm, whose basic feature is a rapid raise to the saturation value, followed by an essentially constant adsorption after it. From the isotherm in Figure 4 one obtains a saturation adsorption value equal to $\Gamma_0 = 1.85$ g NP10 per gram of membrane. Now, it is instructive to ask ourselves not only how much surfactant is adsorbed, but also *how* it adsorbs. As the following figure shall show, the NP10 surfactant adsorbs primarily through the benzene ring (which corresponds to the DPD bead called F, as clearly indicated in figure 2) on most hydrophobic surfaces. Recalling the arguments expressed above, about the potential environmental concerns regarding the release of benzene after the break down of surfactant molecules that include it, such as NP10, it is therefore of paramount importance to ascertain to what extent is presence of benzene indispensable for the performance of the surfactant. This is the key element in the design of a new, environmentally friendly surfactant undertaken in the present work.

Figure 5 shows images obtained from atomistically detailed, microscopic (namely, not DPD) computer simulations of NP10 adsorbed on some typical metal oxide surfaces, carried out using the *Materials Studio* suite of *Accelrys* (see www.accelrys.com). It is important to carry out those studies to confirm the hypothesis that, on entropic or free energy grounds, it is through the benzene ring where NP10 preferentially adsorbs on hydrophobic surfaces, thereby discarding the possibility that it is the loss of atomistic detail what is responsible for the predicted behavior of the surfactant.

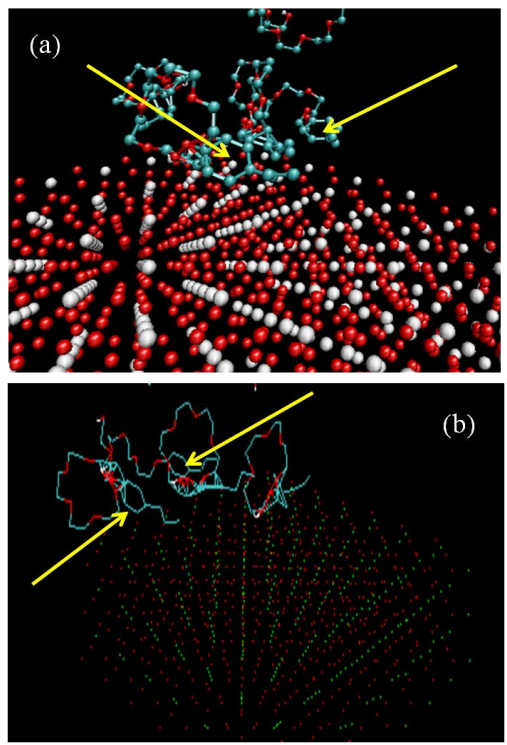

**Figure 5** Adsorption of the surfactant on (a) $Al_2O_3$ surface, and (b) $TiO_2$ surface. The arrows indicate the benzene ring, which is preferentially adsorbed on both surfaces.

Based on the microscopic calculations from which Figure 5 was obtained, as well as those of Figure 3, it is logically sound to speculate as to what the thermodynamic properties of a surfactant such as NP10 *without* the benzene ring would be. If properties such as adsorption isotherms of the newly modified surfactant turn out to be entirely different from those of the proven NP10, then that would be a strong indication that a new strategy ought to be sought for a new surfactant.

The structure of this newly designed surfactant, which shall be called B10 (for "biodegradable surfactant with 10 moles of ethylene oxide) henceforth, is almost the same as that of NP10, with the F – bead (benzene ring) replaced by a G – bead (see Figure 2). It is made up of 12 beads also, with the sequence of beads as follows: H-G-G-G-C-D-D-C-D-D-C-E. The interaction parameters, surfactant concentrations, simulation box volume, interaction with the surfaces, and all other simulation details were chosen exactly as those used in the prediction of the adsorption isotherm of NP10. Figure 6 shows two equilibrium configurations obtained for surfactant B10, one below the micelle formation (left image), and one above it (right image).

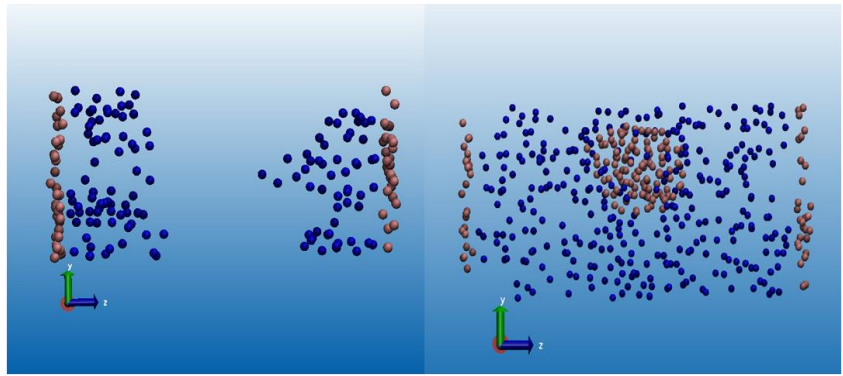

**Figure 6** Adsorption of B10 on model biomembranes. Brown beads represent DPD units named G and H (see mapping in upper panel of Figure 2). Blue beads are the hydrophilic beads, called C, D, and E. The solvent beads have been removed for clarity. The left image corresponds to a B10 concentration [c] = 16 molecules/volume, while the one on the right is for [c] = 40.

By comparing the configurations in Figure 6 for B10 with those in Figure 3 for NP10, it is clear that the same qualitative behaviour is obtained for B10, i.e., the surfactant is driven by two factors. At low enough concentrations, it adsorbs readily on the surfaces, forming brushes. As its concentration is increased, it associates with other B10 molecules, forming micelles, like the one shown on the right in Figure 6, although some of the molecules continue to adsorb on the substrates. With this information I have calculated the adsorption isotherm for B10, which is to be found in Figure 7.

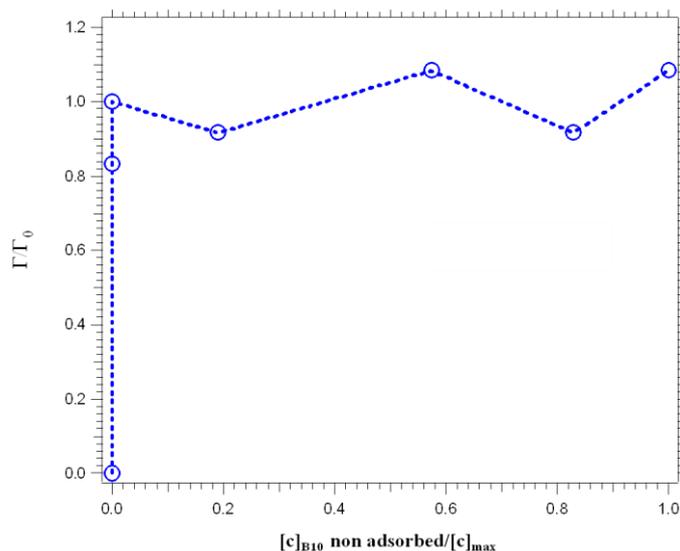

**Figure 7** Adsorption isotherm of surfactant B10, designed in this work. The axes have been normalized to allow for comparisons with other surfactants. The $X$ – axis is the number of non adsorbed B10 molecules, normalized by the maximum amount of non adsorbed surfactant, $[c]_{max}$. The $Y$ – axis represents the number of adsorbed B10 molecules for each concentration added to the dispersion, normalized by the maximum adsorbed amount, at saturation, $\Gamma_0$. The line is only a guide to the eye.

The adsorption isotherm shown in Figure 7 is also of the Langmuir type which, as the images in Figure 6 show, is the result of the fact that B10 adsorbs in monolayers. It should be noted that the competition between adsorption and micelle formation starts before the surfaces have been saturated, and it appears to be driven by the surfactant concentration. The data obtained from the adsorption isotherm allows one to extract the saturation value, which turns out to be $\Gamma_0 = 2.46$ g B10 per gram of membrane. It is slightly larger than that for NP10 (1.85 g), but this has a simple explanation. B10 adsorbs on the surfaces through beads H and G (see Figure 2 for their atomistic mapping), while NP10 does mainly through bead F, of which there is only one per molecule. Other than that, both surfactants have essentially the same thermodynamic behavior, but the newly designed B10 has the bonus that, when used in formulations that interact with the environment, it will biodegrade as easily as NP10, but it will not release benzene freely, because its structure does not include it. It is to be concluded that B10 is an excellent candidate to substitute NP10, given their similar molecular structures, which in turn give rise to very similar function and performance.

## 4 Modelling protein activity through amino acid sequence

The interaction of proteins with biosurfaces is known to be of paramount importance for an ample range of situations (Israelachvili, 2011). More often than not, the specific amino acid sequence determines the activity of the protein in a given environment, even for two proteins of the same molecular weight which are otherwise identical, except for their amino acid sequence. This is a problem that is entirely amenable to be approached by the techniques described here, as shall be shown in what follows. The purpose of this section is to determine the effectiveness of model proteins (meaning that the model does not represent an exact mapping of any given real protein or amino acid) interacting on soft, model biomembranes as a function only their amino acid sequence, within the DPD model. The activity of the proteins with the surfaces is monitored through the calculation of the interfacial tension between them, in an aqueous environment.

Let us start by considering model proteins made up of 24 DPD beads each, with each bead representing a model amino acid. Only three amino acid sequences shall be studied, for brevity. These are shown schematically in Figure 8. All beads have the same size; the molecules are linear, with beads joined by freely rotating harmonic springs of the same type as those used in the previous section, namely with $k_0=100$ and $r_0=0.7$. Soft DPD-like effective surfaces given by equation 5 are placed at the ends of the simulation box in the $z$-direction. Periodic boundary conditions were implemented in the $x$- and $y$-directions, where the fluid is free, but not in the $z$- direction since the walls are impenetrable. The fluid is made up of water monomers and the proteins only. To simulations are performed using a GCMC algorithm, hybridized with DPD, the same used in the previous section. The length of the simulations, time step, box volume, fixed chemical potential, and temperature are exactly the same as used for the prediction of the adsorption isotherms in the previous section.

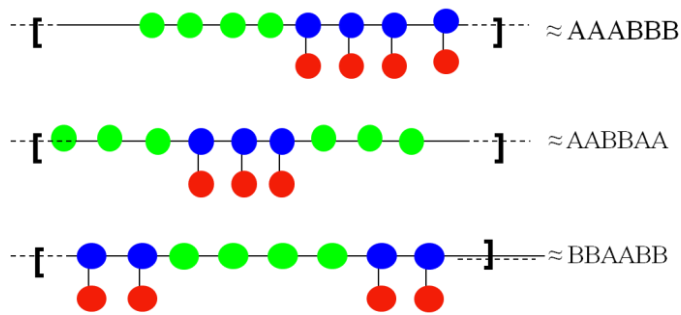

**Figure 8** Three model amino acid sequences for the biomolecules studied in this section. The differently coloured DPD beads are meant to represent different amino acids. The letters on the right symbolize the sequencing, to shorten the notation. The blue – red dimmers are represented in such lettering gas B; however, it is the red bead that one that interacts more strongly with the surfaces. Although all molecules are made up of 24 beads, only 12 are shown in each case for simplicity.

Figure 8 shows schematically the three spatial arrangements, or sequences of the DPD beads; these constitute the only difference between the three cases considered here. The different colouring of the DPD beads is meant to represent the different chemical compositions of the amino acids. The interaction with the surfaces is driven primarily by the red beads as was the case with surfactant NP10 of the previous section. The number of proteins in the simulation box was varied from [c] = 40 – 90 molecules per volume. As for the conservative force constants (see equation 1), they were chosen following the same procedure as in the previous section, and they are given by the matrix shown in Table 2, where the interaction with the walls is listed also.

| $a_{ij}$ | 1 | 2 | 3 | 4 | $a_{iw}$ |
|---|---|---|---|---|---|
| 1 | 78.3 | 78.3 | 171.8 | 155 | 30.75 |
| 2 | | 78.3 | 81.3 | 79.1 | 31.0 |
| 3 | | | 78.3 | 79.1 | 45.0 |
| 4 | | | | 78.3 | 25.0 |

**Table 2** Conservative DPD interaction parameters ($a_{ij}$) used for the beads shown in Figure 8. The numbers identify each type of particle in the simulations, starting with water (1), and the protein DPD beads depicted in Figure 8: green (2), blue

(3), and red (4). The last column lists the interaction ($a_{iw}$) of each type of DPD particle with the surfaces, see equation 5.

Once a sequence among those shown in Figure 8 is chosen, with the appropriate interaction parameters, displayed in Table 2, one proceeds to carry out MC simulations at fixed chemical potential, volume and temperature, at the end of which the components of the pressure tensor, $P_{\alpha\beta}$, are obtained. When equilibrium is reached, these components are calculated using the virial theorem (Allen and Tildesley, 1987). Then, the interfacial interaction between the proteins and the membranes, $\gamma$, is calculated by means of equation 7 (Gama Goicochea et al., 2007):

$$\gamma = L_z \left[ \langle P_{zz} \rangle - \frac{1}{2} \left( \langle P_{xx} \rangle + \langle P_{yy} \rangle \right) \right] \qquad (7)$$

where $<...>$ represents an average over the ensemble, $L_z$ is the length of the simulation box perpendicular to the surfaces, and only the diagonal elements of the pressure tensor are needed. In Figure 9 one finds the final configurations obtained for each sequence.

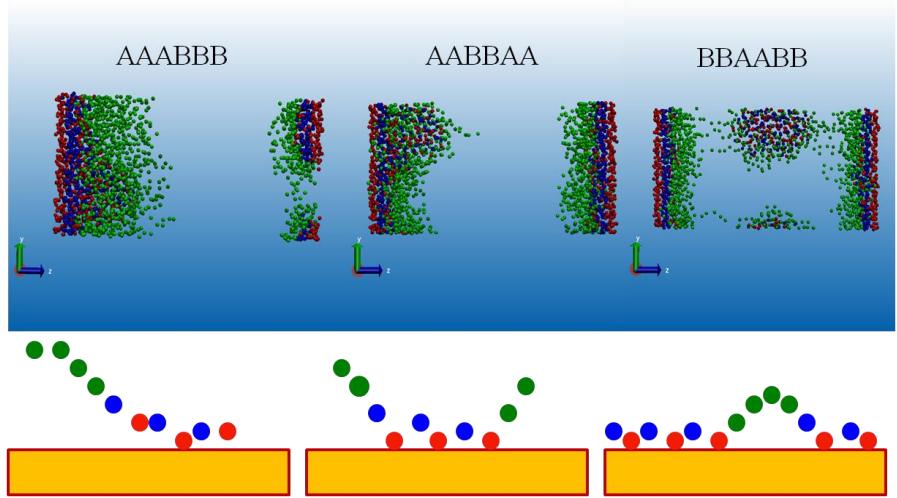

**Figure 9** Interaction of model proteins with membranes. The top part shows the final equilibrium configuration ([c] =90 mol./vol.) for each sequence. The lower is the corresponding schematic of adsorption. The colour code is the same as in Figure 8. Water was removed for simplicity.

The configurations shown in the top part of Figure 9 clearly indicate that, although the number of molecules is same for all three sequences (equal to 90 protein molecules per volume), and although the number of amino acids is equal in all three cases, as well as all other input details, the specific structure of the protein determines its thermodynamic interaction with the surface, and its self association. After proper transformation of DPD units, the following values of the interfacial tension between the proteins and the surfaces are obtained: for the sequence called AAABBB, $\gamma = 44.5 \pm 0.8$ dyn/cm; for AABBAA, $\gamma = 37.5 \pm 0.8$ dyn/cm. Finally, for BBAABB, $\gamma = 59.0 \pm 0.3$ dyn/cm. The conclusion to be drawn from these results is that the most favourable sequence from the thermodynamic point of view between the surfaces and the proteins is given by the one called AABBAA, which corresponds to the image in the middle of Figure 9, with the minimum value of $\gamma$. By contrast, the most unfavourable structure is given by the one labelled BBAABB, because it is the one that requires the most energy investment per unit area, i.e., the largest $\gamma$, when interacting with the surface. From inspection of Figure 9 one sees that the configuration on the far left ("AAABBB") cannot be optimal because it does not cover all the surfaces, which means that the activity of the protein is diminished. On the other hand, the configuration on the far right ("BBAABB") is not the best either, although it appear to adsorb uniformly on the surfaces, because some proteins self – associate in the bulk of the fluid, forming a micelle – like structure, which again means a detriment of the protein function, whose purpose is interacting with the surfaces. The central image in Figure 9 is the best because all proteins are fully interacting with the surfaces through physical adsorption, with none of them in the bulk fluid, which is why the interfacial tension was found to be the minimum for this case. It is to be emphasized that these difference arise purely from a structural difference between the proteins, as laid out by their amino acid sequence.

As for the physical reason behind the results shown in Figure 9 and discussed in the previous paragraph, the caricature adsorption model shown in the lower images in Figure 9 gives us some insight. The structure with the minimum interfacial tension ("AABBAA") has the particular feature that it groups together as nearest neighbours, all the amino acids that preferentially adsorb on the surfaces (shown in red in Figure 8). This maximises the area covered on the surface by the molecule. If the red amino acid were surrounded by green or blue ones as nearest neighbours, which do not adsorb as favourable as red on the surfaces (as is the case for structures

AAABBB and BBAABB) then the energy cost per unit area of the protein – surface interaction turns out to be more expensive, and it is therefore not preferred by thermodynamics. These results and their interpretation are in complete agreement with those of recent experimental studies (Jhon, Y. K. et al., 2009).

## 5 Conclusions

The present work reports the design and thermodynamic testing of a biodegradable surfactant and model proteins in aqueous solutions confined by soft surfaces, intended to model biologically relevant membranes, by means of mesoscopic DPD computer simulations. In the first part of this report it was shown that a newly designed linear, non ionic and environmentally friendly surfactant performs at least as well as its benzene - containing (and therefore, *not* environmentally friendly) counterpart. The performance of both surfactants was tested with the calculation of adsorption isotherms and surface saturation. The new surfactant structure was proposed after a careful microscopic analysis of the role played by each part of the original surfactant's structure. The second part of this research was devoted to the design of differently sequenced but otherwise equal model proteins, with the purpose of determining which one had the optimal activity when interacting with a biologically important surface. Predictions of their interfacial (protein – surface) tension led to the conclusion that the proteins that adsorbed uniformly on the surfaces without leaving any of them in the bulk to form energy – consuming micelles were the ones preferred on thermodynamic grounds. These conclusions were found to be fully supported by recent experiments reported in the literature.

The advantages of carrying out coarse – grained, mesoscopic DPD simulations like the ones reported here are numerous. Not only can one reach length scales of the order of μm and times scales of ms, but because of the momentum conserving structure of the DPD forces, one can capture complex mesoscopic hydrodynamic behaviour that is crucial for the study of the association of biomolecules. Other advantages include the fact that simulations can be carried out with many chains at once, and with the solvent included explicitly. The latter is critical to incorporate excluded volume interactions at short distances.

The two cases studied in this work have a common thread, which can perhaps be summarized as follows: when comparing the performance of two molecules whose only difference is their structure, then any difference in their thermodynamic performance must be attributed precisely to their structural dissimilarities. This has been more eloquently stated by Crick, in his famous dictum "*if you want to understand function, study structure*" (Crick, 1988). These studies can be considered as a stepping stone toward the construction of more atomistically detailed, albeit more computationally expensive models.

## 7 Acknowledgments

The author wishes to thank M. Maciel, S. Viale, F. Zaldo, and especially N. López for enlightening discussions during the early stage of this project. Additional insightful comments from G. Pérez Hernandez are gratefully acknowledged. This work was initially sponsored by the Centro de Investigación en Polímeros (CIP, COMEX Group) and subsequently by PROMEP, Project 47310286-912025.